\documentclass{ws-p8-50x6-00}

\def\rightnote{{\it Presented at {\bf EMI 2001}, Osaka, Japan}}%

\begin{document}

\thispagestyle{empty}
\begin{flushright}  
\setlength{\baselineskip}{2.6ex}
BU-HEPP-06-01\\
TRI-PP-02-01\\
\end{flushright}
\vspace{1.0cm}

\begin{center}
{\huge  \bf Strange-quark Current in the Nucleon from Lattice QCD}\\
\vspace{0.8cm}
{Randy Lewis}\\ 
\vspace{0.3cm}
{\small\em Department of Physics, University of Regina,Regina SK, Canada S4S 0A2\\}
\vspace{0.6cm}
{W. Wilcox}\\ 
\vspace{0.3cm}
{\small\em Department of Physics, Baylor University, Waco TX 76798-7316, U.S.A.\\}
\vspace{0.6cm}
{R.M. Woloshyn}\\ 
\vspace{0.3cm}
{\small\em TRIUMF, 4004 Wesbrook Mall, Vancouver BC, Canada V6T 2A3\\}
\end{center}
\vspace{0.8cm}
\noindent
Presented at the International Symposium on Electromagentic Interactions
in Nuclear and Hadron Physics, Osaka, Japan, 4th to 7th December, 2001.
\newpage
\thispagestyle{empty}
\
\newpage

\title{Strange-quark Current in the Nucleon from Lattice QCD}

\author{Randy Lewis}

\address{Department of Physics, University of Regina, Regina SK, Canada S4S 0A2}

\author{W. Wilcox}

\address{Department of Physics, Baylor University, Waco TX 76798-7316, U.S.A.}

\author{R.M. Woloshyn}

\address{TRIUMF, 4004 Wesbrook Mall, Vancouver BC, Canada V6T 2A3}


\maketitle

\abstracts{
The contribution of the strange-quark current to the electromagnetic form
factors of the nucleon is studied using lattice QCD.
The strange current matrix elements from our lattice calculation are analyzed
in two different ways, the differential method used in an earlier work
by Wilcox
and a cumulative method which sums over all current insertion times.
The preliminary results of our simulation indicate the importance of high
statistics, and that consistent results between the varying analysis methods
can be achieved. Although this simulation does not yet yield a number that 
can be compared to experiment, several criteria useful in assessing
the robustness of a signal extracted from a noisy background are presented.
}

\section{Introduction}
An important theme of contemporary Hadron Physics is the role of nonvalence
degrees of freedom. In particular, the contribution of strange quarks to a 
variety of nucleon properties has been studied.\cite{Liu:2001ei}
For nucleon form factors the 
contribution of the strange-quark current can be extracted using information
obtained from parity violation in polarized electron 
scattering.\cite{Beck:2001yx} A number of
experimental results have been reported\cite{Hasty:2000ep,Aniol:2001at}
and new measurements are planned.
As well, there are numerous calculations using a number of different 
approaches.\cite{Beck:2001yx}

In this work we focus on the calculation of the strange-quark current loop
using lattice QCD. Calculations of these so-called disconnected current
insertions have been reported 
previously.\cite{Wilcox:2001qa,Dong:1998xr,Mathur:2001cf}
However, these calculations differ
in the method used to analyze the results of their lattice simulations and
also differ in their conclusions. Wilcox\cite{Wilcox:2001qa}
used a differential method to analyze
the time dependence of the three-point function and no signal for the strange
current was found. On the other hand, 
Dong and co-workers \cite{Dong:1998xr,Mathur:2001cf}
used a summation of 
the current insertion over lattice times which requires a further 
identification and fitting of a linear time dependence. They claim to see a 
definite signal for the strangeness form factors.

\begin{figure}[t]
\begin{center}
\epsfxsize=22pc 
\epsfbox{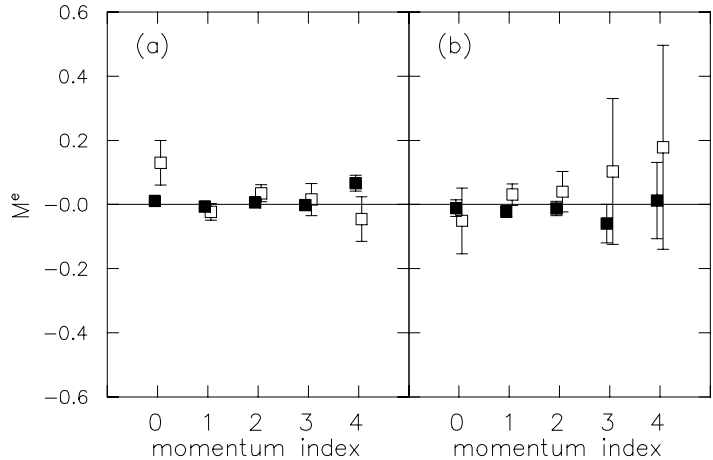} 
\caption{Comparison of the electric matrix element calculated with the
differential method for samples of 100 configurations (open symbols) and
1050 configurations (filled symbols). (a) Average of times 10 to 12.
(b) Average of times 15 to 17.  \label{fig:emi01}}
\end{center}
\end{figure}

We present preliminary results from a lattice QCD simulation of the 
strange-quark current loop using a Monte-Carlo sample of gauge field
configurations about ten times larger than in previous studies. 
Both differential and cumulative time analyses are carried out on the
simulation data. Comparisons of results obtained from a 100 configuration
subsample (the same size used in previous 
work\cite{Wilcox:2001qa,Dong:1998xr,Mathur:2001cf}) with results from the full 
data set show quite clearly how large fluctuations, that could be interpreted 
as a signal in low-statistics data, disappear with improved statistics.
Using our complete data set consistency between different analysis methods
is achieved. By comparing different analysis methods and results using
different sized gauge field samples, criteria for assessing the robustness
of a strange quark current signal are presented.
Using these criteria, no compelling signal for the strange quark current 
is observed. 

\section{Lattice  Calculations}
The standard methods of the path integral formulation of lattice QCD in
Euclidean space are used. The Wilson action is used for both quark and
gluon fields. We need two and three-point functions which describe,
respectively, the propagation of nucleon states as a function of Euclidean
lattice time and the strange quark current in the presence of a nucleon.
The two-point function $G^{(2)}(t;\overrightarrow{q})$ correlates the excitation
of a nucleon state with momentum $\overrightarrow{q}$ 
at some initial time (called 0) and its annihilation at time
t. For large Euclidean time t this quantity decreases exponentially 
like $e^{-E_{q}t}$.

\begin{figure}[t]
\begin{center}
\epsfig{file=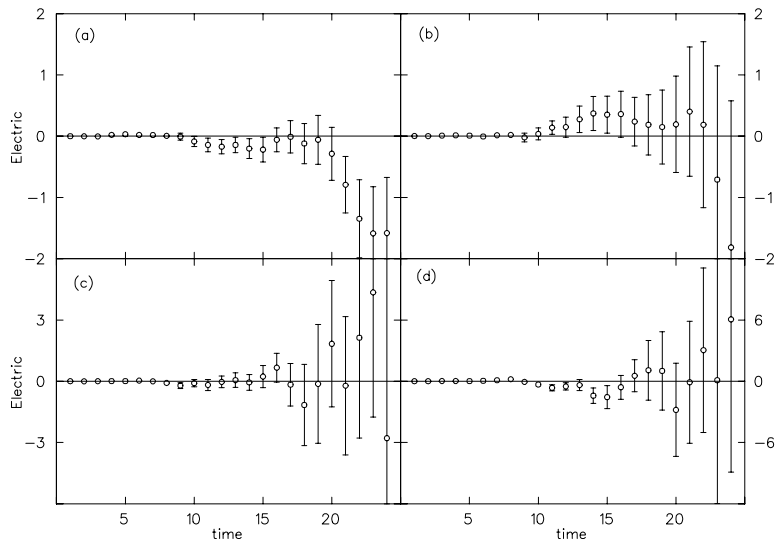,angle=90,width=24pc}
\caption{Electric matrix element using the
cumulative method for a sample of 100 configurations. 
The plots correspond to momentum transfer (a) (1,0,0), (b) (1,1,0),
(c) (1,1,1), (d) (2,0,0).
\label{fig:emi02}}
\end{center}
\end{figure}

\begin{figure}[t]
\begin{center}
\epsfig{file=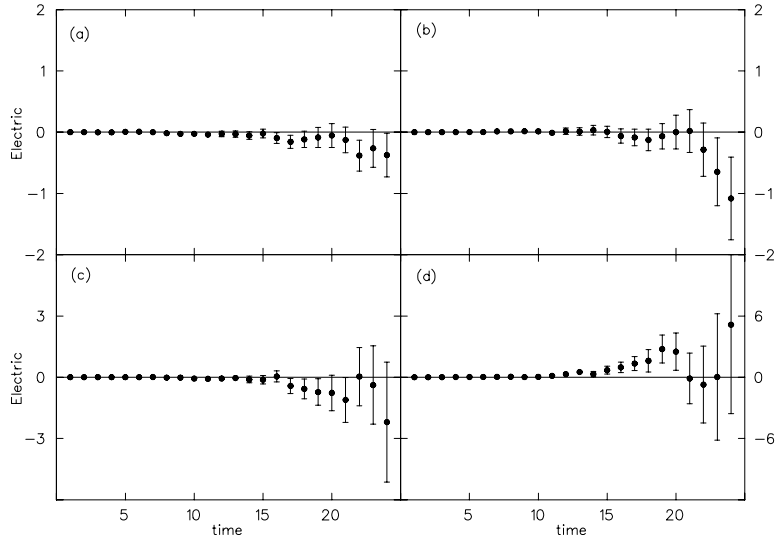,angle=90,width=24pc}
\caption{Electric matrix element using the
cumulative method for a sample of 1050 configurations. 
The plots correspond to momentum transfer (a) (1,0,0), (b) (1,1,0),
(c) (1,1,1), (d) (2,0,0).
\label{fig:emi03}}
\end{center}
\end{figure}

To calculate the three-point function $G^{(3)}(t,t';\overrightarrow{q})$ an
insertion of the strange-quark vector current is made at time $t'$. Since there
are no strange valence quarks in the nucleon this insertion amounts to a 
correlation of a strange-quark current loop with the nucleon two-point function.
The strange-quark current loop is calculated using a so-called noisy 
estimator with $Z_2$ noise\cite{Dong:xb,Dong:1993pk}.
The noise and perturbative subtraction methods employed here are the 
same as those of Wilcox\cite{Wilcox:2001qa}
except that 60 noises are used instead of 30.

The three-point function
also has an exponential time dependence but is more complicated
due to the presence of two times. It can be shown that the exponential time 
factors can be cancelled by taking the ratio 

\begin{equation}
R(t,t';\overrightarrow{q}) = \frac{G^{(3)}(t,t';\overrightarrow{q})G^{(2)}(t';0)}
{G^{(2)}(t;0)G^{(2)}(t';\overrightarrow{q})}.
\label{eq:ratio}
\end{equation}
It is these ratios R that are analyzed to get the final results.

Note that in writing the two and three-point functions, labels associated with 
the  Dirac indices of the nucleon fields and the Lorentz index of the current
have been suppressed. Also in writing Eq.~(\ref{eq:ratio}) the necessary spin 
sums and projections are not shown, the expression is given schematically to
show the time dependence of the various factors. The detailed calculations
follow previous work.\cite{Wilcox:2001qa,Dong:1998xr,Mathur:2001cf}

The ratio R is usually summed in order to try to improve the signal. The
differential method uses a difference of $R(t,t';\overrightarrow{q})$ on 
neighbouring time slices. It can be shown\cite{Wilcox:2001qa}
that the quantity 
$ M(\overline{t},\overrightarrow{q})$ given by
\begin{equation}
M(\overline{t},\overrightarrow{q})=\sum ^{t+1}_{t'=1}(R(t,t';\overrightarrow{q})
-R(t-1,t';\overrightarrow{q}))
\label{eq:diff}
\end{equation}
is the current matrix element of interest.

\begin{figure}[t]
\begin{center}
\epsfxsize=22pc 
\epsfbox{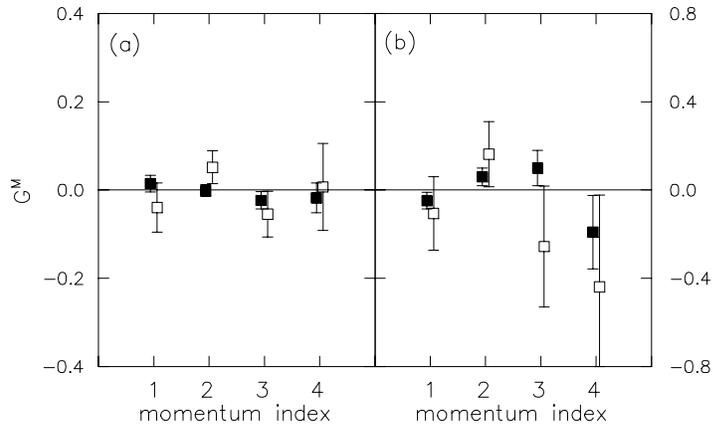} 
\caption{Comparison of the magnetic matrix element calculated with the
differential method for samples of 100 configurations (open symbols) and
1050 configurations (filled symbols) using $\kappa=0.152$ for the valence
quark. (a) Average of times 10 to 12.
(b) Average of times 15 to 17.  \label{fig:emi04}}
\end{center}
\end{figure}

\begin{figure}[t]
\begin{center}
\epsfxsize=22pc 
\epsfbox{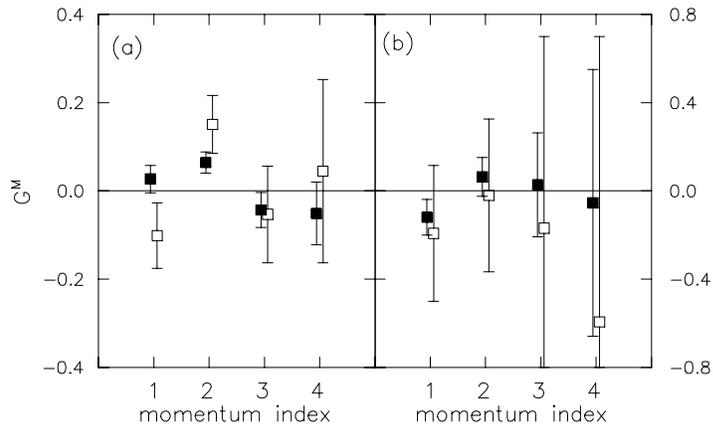} 
\caption{Comparison of the magnetic matrix element calculated with the
differential method for samples of 100 configurations (open symbols) and
1050 configurations (filled symbols) using $\kappa=0.154$ for the valence
quark. (a) Average of times 10 to 12.
(b) Average of times 15 to 17.  \label{fig:emi05}}
\end{center}
\end{figure}

An alternative is to simply sum R and then fit to a linear time dependence to
get the matrix element. We use 
\bea
S(t,\overrightarrow{q}) & = & \sum ^{t'=t}_{t'=1}R(t,t';\overrightarrow{q}),\\
 & \rightarrow  & constant+tM(t,\overrightarrow{q}).
\label{eq:sum}
\eea
Summing current insertions up to $t'=t$ follows the suggestion of Viehoff
{\em et~al.}\cite{Viehoff:1998wi} and helps to reduce the statistical noise. 
Dong and co-workers\cite{Dong:1998xr,Mathur:2001cf}
actually use a different upper limit ($t'=t_{fixed}, t_{fixed}>t$). 

\section{Results}

\begin{figure}[t]
\begin{center}
\epsfig{file=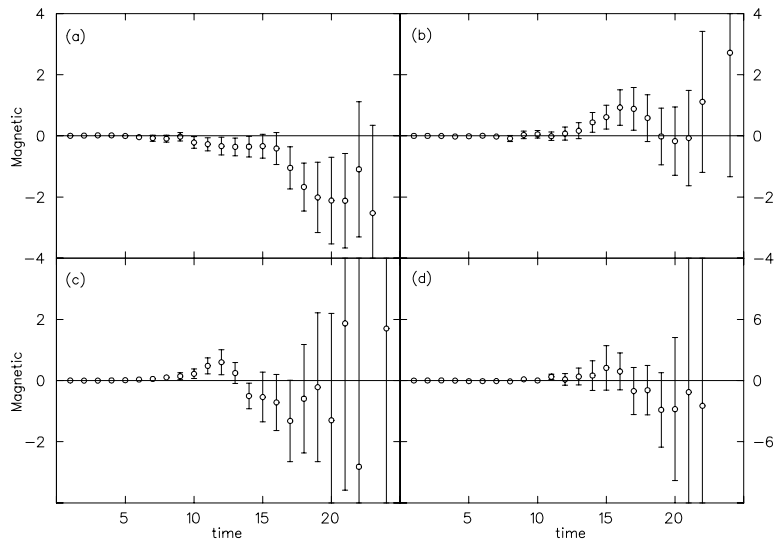,angle=90,width=24pc}
\caption{Magnetic matrix element using the
cumulative method for a sample of 100 configurations with  $\kappa=0.152$
valence quark. The plots correspond to momentum transfer 
(a) (1,0,0), (b) (1,1,0), (c) (1,1,1), (d) (2,0,0).
\label{fig:emi06}}
\end{center}
\end{figure}

\begin{figure}[t]
\begin{center}
\epsfig{file=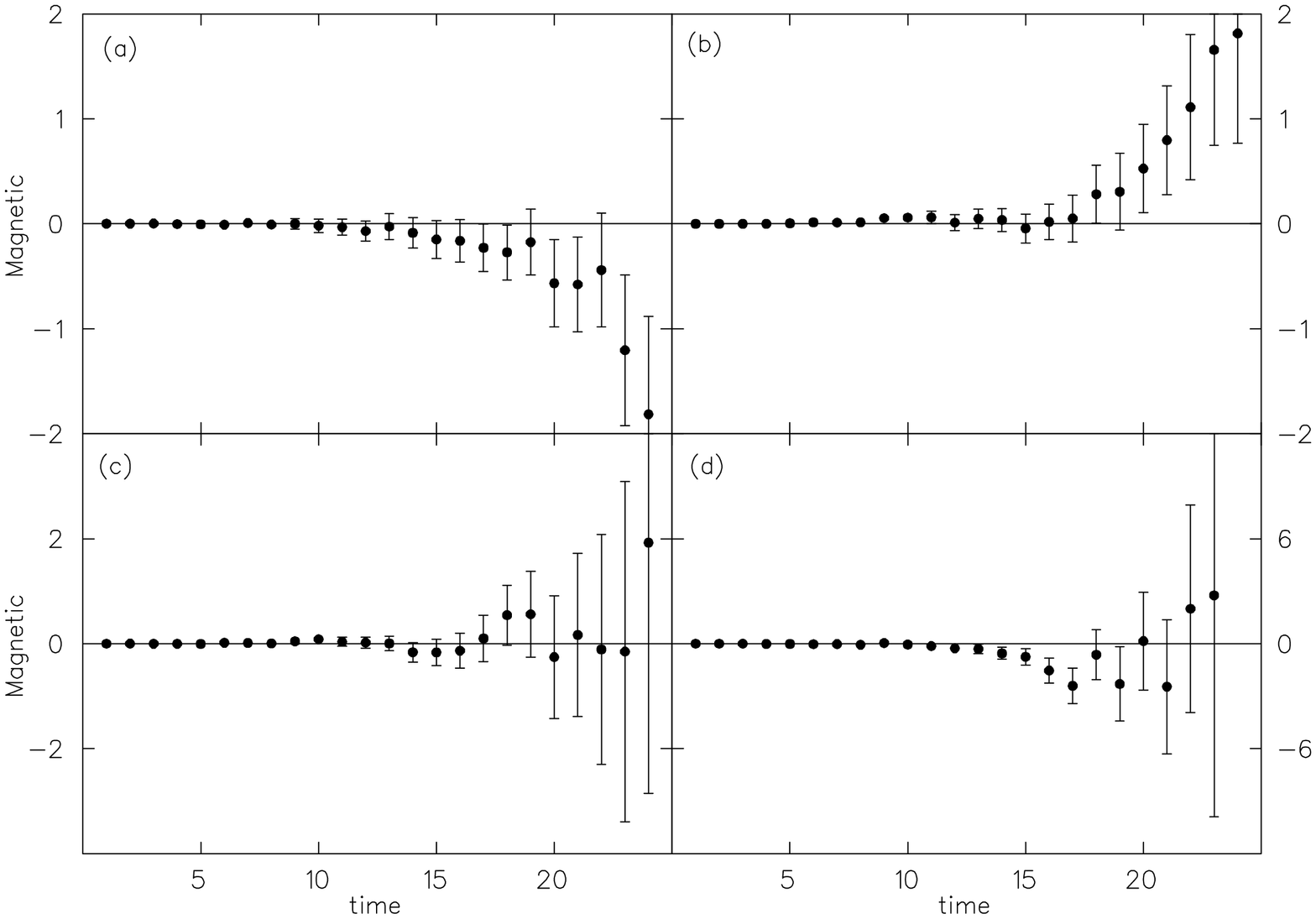,angle=90,width=24pc}
\caption{Magnetic matrix element using the
cumulative method for a sample of 1050 configurations with  $\kappa=0.152$
valence quark. The plots correspond to momentum transfer 
(a) (1,0,0), (b) (1,1,0), (c) (1,1,1), (d) (2,0,0).
\label{fig:emi07}}
\end{center}
\end{figure}

\begin{figure}[t]
\begin{center}
\epsfig{file=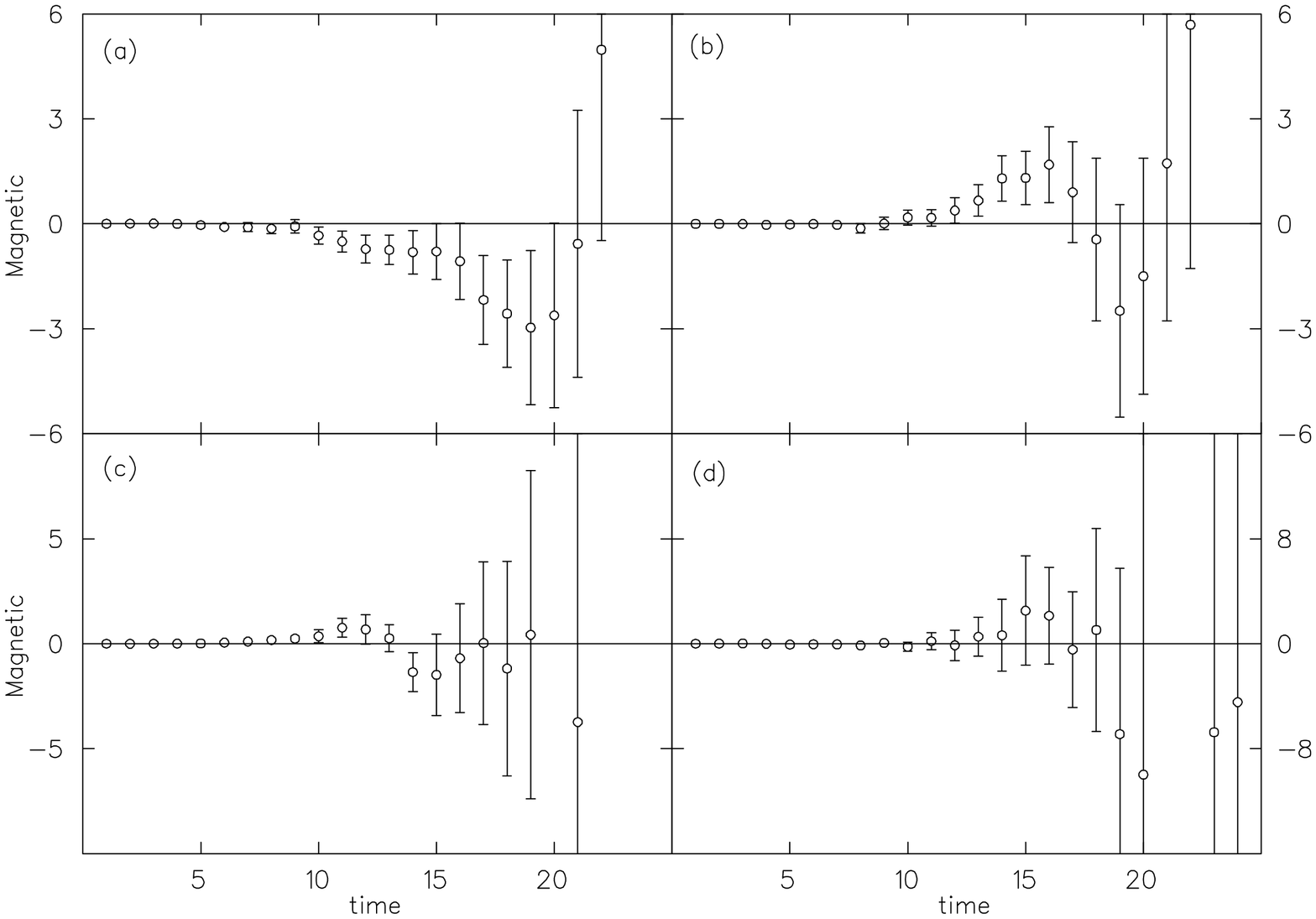,angle=90,width=24pc}
\caption{Magnetic matrix element using the
cumulative method for a sample of 100 configurations with  $\kappa=0.154$
valence quark. The plots correspond to momentum transfer 
(a) (1,0,0), (b) (1,1,0), (c) (1,1,1), (d) (2,0,0).
\label{fig:emi08}}
\end{center}
\end{figure}

\begin{figure}[t]
\begin{center}
\epsfig{file=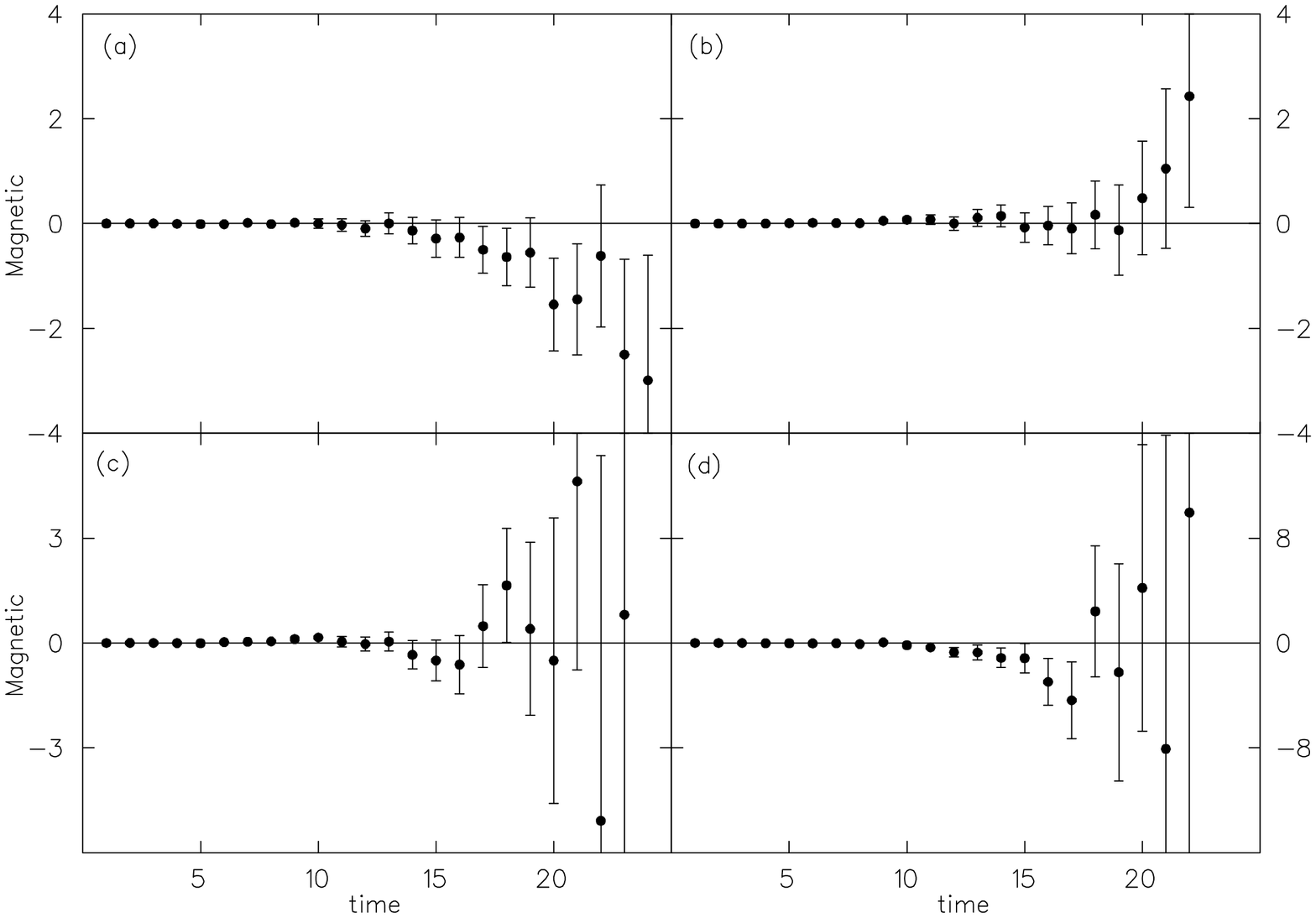,angle=90,width=24pc}
\caption{Magnetic matrix element using the
cumulative method for a sample of 1050 configurations with  $\kappa=0.154$
valence quark. The plots correspond to momentum transfer 
(a) (1,0,0), (b) (1,1,0), (c) (1,1,1), (d) (2,0,0).
\label{fig:emi09}}
\end{center}
\end{figure}

Calculations were carried out in quenched approximation at gauge field coupling
of $\beta = 6.0$. The lattice size was $20^3 \times 32$. A total of 1050 gauge
field configurations were generated using a pseudo-heat bath algorithm. Two
thousand sweeps were used between saved configurations.

Since quenched lattice QCD does not provide a perfect description of hadrons,
there is some ambiguity in fixing the parameters (overall scale and quark masses)
in the calculations. In this work we use 0.152 as the hopping parameter for the
strange quark. Using $a^{-1}=2GeV$ and the $\phi$-meson to fix the strange quark
mass would suggest a hopping parameter closer to $\kappa_s=$0.153. On the other
hand, using the scale of Dong and co-workers\cite{Dong:1998xr,Mathur:2001cf}
$a^{-1}=1.74GeV$ gives $\kappa_s$ smaller than 0.152.

Results for electric matrix element with valence quark $\kappa_v=$0.152
calculated using the differential method
Eq.~(\ref{eq:diff}) are shown in Fig.~(\ref{fig:emi01}), averaging over two
different time windows. The 100 configuration sample results are consistent
with those obtained by Wilcox\cite{Wilcox:2001qa}. No signal is seen when
the gauge field sample size is increased to 1050. The summed ratio Eq.~(3)
is shown in Fig.~(\ref{fig:emi02}) and (\ref{fig:emi03}) as a function of
the nucleon sink time for different momentum transfers. It shows
oscillations characteristic of lattice correlation function 
ratios\cite{Aoki:1996bb}.
The magnitude of these oscillations decreases slowly as the configuration 
sample size is increased.

Results for the magnetic matrix element for different valence quark masses
calculated using the differential method are plotted in 
Fig.~(\ref{fig:emi04}) and (\ref{fig:emi05}).  As in the electric case 
the results are consistent with zero.

Finally, the summed ratios for the magnetic current are given in 
Fig.~(\ref{fig:emi06}) to Fig.~(\ref{fig:emi09}). Note that a kinematic
factor (see Eq.~(3) in Ref.\cite{Dong:1998xr}) 
of $q/(E+M)$ has been removed. The results for 100 configurations, 
Fig.~(\ref{fig:emi06}) and (\ref{fig:emi08}), should
be compared to Fig.~(1) of Dong {\em et al.}\cite{Dong:1998xr} and
Fig.~(2) of Mathur and Dong\cite{Mathur:2001cf} where calculations with
the same statistics are reported. Then, comparing with 
Fig.~(\ref{fig:emi07}) and (\ref{fig:emi09}), one sees that the kind of
oscillations in the time range 10 to 15 which
Dong and co-workers\cite{Dong:1998xr,Mathur:2001cf} took 
to be their signal, have largely disappeared in the higher statistics 
results. Of course, fluctuations still persist at larger times but even
higher statistics simulations will be necessary to establish if they go away
or if indeed a strange quark current signal is hiding under them.

\section{Conclusions}

To get an estimate of the strange-quark current matrix elements requires the
extraction of a small signal in the presence of large statistical fluctuations.
The results presented here suggest a number of useful criteria that should be 
met before one can claim a credible signal:
\begin{itemize}
\item There should be consistency between different analysis methods.
\item The signal should appear in the same lattice time region and its
statistical significance should increase as the size of the Monte-Carlo 
sample of gauge fields is increased.
\item The signal should appear in the same time window for different masses.
\end{itemize}

Not all of these criteria have been met at our present level of statistics.
Work is continuing and it is hoped to have final results with an increased
configuration sample size in the not too distant future.

\section*{Acknowledgments}
This is work is supported in part by the National Science Foundation
under grant 0070836 and in part by the Natural Sciences
and Engineering Research Council of Canada. We thank N. Mathur for 
helpful discussions.

\end{document}